\documentclass[a4paper]{article}

\usepackage{INTERSPEECH2021}

\usepackage{anyfontsize}
\usepackage{t1enc}

\usepackage{graphicx}
\usepackage{amssymb,amsmath,bm}
\usepackage{textcomp}
\usepackage{color}
\usepackage{multirow}
\usepackage{hyperref}
\usepackage{url}
\usepackage[activate]{microtype}
\usepackage{subcaption}
\usepackage{soul}
\usepackage{cleveref}
\usepackage[activate]{microtype}
\usepackage{ulem}
\usepackage[nolist]{acronym}

\newcommand{\eg}{e.\,g., }
\newcommand{\ie}{i.\,e., }

\newcommand{\et}{{et al.\,}}

\def\vec#1{\ensuremath{\bm{{#1}}}}

\newcommand{\ds}{\mbox{\textsc{Deep Spectrum}}}

\PassOptionsToPackage{dvipsnames,table}{xcolor}
\usepackage{pgfplotstable}
\pgfplotsset{compat=1.14}

\ninept

\title{On the Impact of Word Error Rate on Acoustic-Linguistic Speech Emotion Recognition: An Update for the Deep Learning Era}
\name{Shahin Amiriparian$^{1}$, Artem Sokolov$^{2,3}$, Ilhan Aslan$^{2}$, Lukas Christ$^{1}$, Maurice Gerczuk$^{1}$, Tobias~H\"ubner$^{1}$, Dmitry Lamanov$^{2}$, Manuel Milling$^{1}$, Sandra Ottl$^{1}$, Ilya Poduremennykh$^{2}$,  Evgeniy~Shuranov$^{2,4}$, Bj\"orn W.\ Schuller$^{1,5}$}

\address{\fontsize{11}{11}\selectfont 
  $^1$EIHW -- Chair of Embedded Intelligence for Health Care and Wellbeing, University of Augsburg, Germany\\
  $^2$Huawei Technologies \\
  $^3$HSE University, Nizhniy Novgorod, Russia  \\
  $^4$ITMO University, Saint Petersburg, Russia \\
  $^5$GLAM -- Group on Language, Audio, \& Music, Imperial College London, UK}
\email{shahin.amiriparian@uni-a.de, sokolov.artem@huawei.com, ilhan.aslan@huawei.com, lukas.christ@informatik.uni-augsburg.de, maurice.gerczuk@informatik.uni-augsburg.de, tobias.huebner@informatik.uni-augsburg.de, lamanov.dmitry@huawei.com, manuel.milling@informatik.uni-augsburg.de, sandra.ottl@informatik.uni-augsburg.de, poduremennykh.ilya@huawei.com, evgeniy.shuranov@huawei.com, schuller@ieee.org}

\begin{document}

\maketitle
\begin{abstract}
Text encodings from automatic speech recognition (ASR) transcripts and audio representations have shown promise in speech emotion recognition (SER) ever since. Yet, it is challenging to explain the effect of each 
information stream
on the SER systems. Further, more clarification is required for analysing the impact of ASR's word error rate (WER) on linguistic emotion recognition per se and in the context of fusion with acoustic information exploitation in the age of deep ASR systems.
In order to tackle the above issues we create transcripts from the original speech by applying three modern ASR systems, including an end-to-end model trained with recurrent neural network-transducer loss, a model with connectionist temporal classification loss, and a \textsc{wav2vec} framework for self-supervised learning. 
Afterwards, we use pre-trained textual models to extract text representations from the ASR outputs and the gold standard. For extraction and learning of acoustic speech features, we utilise \textsc{openSMILE}, \textsc{openXBoW}, \textsc{DeepSpectrum}, and \textsc{auDeep}. Finally, we conduct decision-level
fusion on both information streams -- acoustics and linguistics.
Using the best development configuration, we achieve state-of-the-art unweighted average recall values of $73.6\,\%$ and $73.8\,\%$ on the speaker-independent development and test partitions of IEMOCAP, respectively.

\end{abstract}
\noindent\textbf{Index Terms}: emotion recognition, automatic speech recognition, computational paralinguistics

\section{Introduction}
\label{sec:introduction}

As technology is becoming increasingly ubiquitous, speech input is gaining popularity as an accessible interaction modality. 
The rise of voice assistants, \eg Amazon's Alexa, exemplifies this trend.
While today's technologies may understand speech commands
well, the conversation quality is still far from what we as humans experience in interpersonal communication. Emotional expressions are a key part of interpersonal communication.
 They are embodied in our gestures, body posture, and 
speech. 
Humans typically express and recognise emotional speech effortlessly,
while, for machines, recognising emotions in speech is still a hard challenge.

In this paper, we present 
an update to previous research (\ie \cite{metze2010}) 
on the trade-off between \ac{ASR} accuracy (\ie \ac{WER}) and linguistic emotion recognition, and the impact thereof on the later fusion with voice based emotion recognition. 
Such an update is urgently required, as I) most papers analysing the fusion of acoustics and linguistics use human transcripts and not actual ASR (\eg ~\cite{cambria2017benchmarking},~\cite{cho2019deep},~\cite{chen2020multi}), hence, oversimplifying the problem,
and II) practically no systematic investigation of the WER on linguistic \ac{SER} exists, besides \cite{metze2010} -- however, more than a decade since this investigation has witnessed massive improvements in ASR in the era of deep ASR approaches, and III) the modelling of linguistic information itself has changed dramatically with the advent of deep text modelling and the existence of large pre-trained according models. Hence, a re-investigation is urgently needed.
To provide a comprehensive and ecologically valuable overview, we juxtapose and contrast variations of contemporary solutions for \ac{ASR} and feature sets for emotion recognition from both text and voice. We highlight the best performing fusion solution, that to 
the best of our knowledge sets a new state-of-the-art. Further, we describe in detail the overall system that we use for our experiments. Moreover, we provide different variations of every system's component. 

\section{Methodology}
\label{sec:systems}
In this section, we introduce feature extraction methods that are well suited to process acoustic and linguistic cues. The features are used as inputs to \acp{SVM} and, therefore, build the basis for our \ac{SER} analysis. We further introduce several \ac{ASR} approaches, which will be investigated with respect to their \ac{WER} and corresponding suitability in the \ac{SER} context. 

\subsection{Audio Features} 
\label{ssec:audio}
We examine four different audio feature sets. The first feature set is extracted with the \textsc{openSMILE} toolkit using the ComParE\_2016.conf configuration file~\cite{Eyben13-RDI}. It contains $6\,373$ static features resulting from the computation of functionals (statistics) over \ac{LLD} contours\footnote{\href{https://github.com/audeering/opensmile}{https://github.com/audeering/opensmile}}~\cite{Eyben13-RDI,Schuller13-TI2}.
A full description of the feature set can be found in \cite{Weninger13-OTA}.

In addition to the default \ac{ComParE} feature set, we provide \ac{BoAW} features by using \textsc{openXBOW}~\cite{Schmitt17-OIT}.
These have been applied successfully for, \eg acoustic event detection~\cite{Lim15-RSE} and speech-based emotion recognition~\cite{Schmitt16-ATB}.
After a quantisation based on a codebook, audio chunks are represented as histograms of acoustic \acp{LLD}. One codebook is learnt for $65$ \acp{LLD} from the \textsc{ComParE} feature set, and another one for $65$ deltas of these \acp{LLD}.
Codebook generation is done by \textit{random sampling} from the \acp{LLD} and its deltas in the training data. Each \ac{LLD} and delta is assigned to $10$ audio words from the codebooks with the lowest Euclidean distance. Subsequently, both \ac{BoAW} representations are concatenated. Finally, a logarithmic term frequency weighting is applied to compress the numeric range of the histograms.

The feature extraction \ds{} toolkit\footnote{\href{https://github.com/DeepSpectrum/DeepSpectrum}{https://github.com/DeepSpectrum/DeepSpectrum}} is applied to obtain deep representations from the input audio data utilising pre-trained \acp{CNN}~\cite{Amiriparian17-SSC}. \ds{} features have been shown to be effective, \eg for speech processing~\cite{amiriparian2019deep,Amiriparian18-BND} and sentiment analysis~\cite{Amiriparian17-SAU}.
First, audio signals are transformed into Mel-spectrogram plots using a Hanning window of width $32$\,ms and an overlap of $16$\,ms. From these, $128$ Mel frequency bands are computed. The spectrograms are then forwarded through a pre-trained \textsc{DenseNet121}~\cite{huang2017densely} and the activations from the `avg\_pool' layer are extracted, resulting in a $1\,024$ dimensional feature vector. 

Another feature set is obtained through unsupervised representation learning with recurrent sequence-to-sequence autoencoders, using \textsc{auDeep}\footnote{\href{https://github.com/auDeep/auDeep}{https://github.com/auDeep/auDeep}} \cite{Amiriparian17-STS,Freitag18-AUL}. 
This feature set models the inherently sequential nature of audio with \acp{RNN} within the encoder and decoder networks~\cite{Amiriparian17-STS,Freitag18-AUL}.
First, Mel-scale spectrograms are extracted from the raw waveforms.
In order to eliminate some background noise, power levels are clipped below four different given thresholds in these spectrograms. The number of thresholds results in four separate sets of spectrograms per data set. Subsequently, a distinct recurrent sequence-to-sequence autoencoder is trained on each of these sets of spectrograms in an unsupervised way, \ie without any label information. The learnt representations of a spectrogram are then extracted as feature vectors for the corresponding instance. Finally, these feature vectors are concatenated to obtain the final feature vector. For the results shown in~\Cref{tab:Baselines}, the autoencoders' hyperparameters are not fine tuned.

\subsection{Text Features}
\label{ssec:text_baselines}

DeepMoji, proposed by Felbo \et~\cite{felbo2017using}, is a model pre-trained for emotion-related text classification tasks. It consists of two bidirectional \ac{LSTM} layers, followed by an attention layer and yields a sentence encoding of length $2\,304$. 
Even though DeepMoji is pre-trained on emotional tweets only, the authors show that it also performs well for other kinds of emotional text data, \eg reports of emotional experiences.  
We extract DeepMoji sentence encodings via the PyTorch implementation \textit{TorchMoji}\footnote{\href{https://github.com/huggingface/torchMoji}{https://github.com/huggingface/torchMoji}}.

Moreover, we employ several variants of \ac{BERT}~\cite{devlin2018bert} that has set new standards for many text processing tasks in recent years. In its \textit{base} configuration, BERT consists of $12$ transformer (\cite{vaswani2017attention}) encoder layers. This network is pre-trained on large text data sets using two unsupervised language modelling tasks, namely masked word prediction and next sentence prediction. Here, we employ the pre-trained BERT-base model to obtain sentence encodings. 
The output of the last layer's hidden state for the special token [CLS], followed by one tanh-activated linear layer (\textit{pooler\_output}) is considered as the sentence encoding.

ALBERT (\textit{A Lite BERT})~\cite{lan2019albert} is a popular variant of BERT. It is of the same size as the original BERT model but comes with considerably less parameters due to parameter sharing across layers and factorisation of the embedding matrix. Furthermore, the next sentence prediction task in BERT's pre-training has been replaced by sentence order prediction, \ie deciding whether two sentences are given in the correct order. ALBERT has been shown to outperform BERT on many tasks. Similar to our BERT baseline, we take
the \textit{pooler\_output} of ALBERT in its \textit{base} version as our sentence encoding.

Another recent variant of the BERT \ac{LM} is given by ELECTRA~\cite{clark2020electra}, referring to an alternative method of pre-training transformer language models. In this approach, corrupted input words are detected. First, a generator model corrupts the input sentence. Then, the discriminator, \ie the actual language model, predicts for every word whether it has been changed by the generator or not. BERT-like transformer networks pre-trained in this fashion outperform other BERT variants on several tasks. The architecture of the model is nearly identical to BERT-base. We take the embedding of the special token [CLS] as the sentence encoding.
For all three BERT variants, huggingface implementations and pre-trained weights \footnote{\href{https://huggingface.co/bert-base-cased}{https://huggingface.co/bert-base-cased}} \footnote{\href{https://huggingface.co/albert-base-v2}{https://huggingface.co/albert-base-v2}} \footnote{\href{https://huggingface.co/google/electra-base-discriminator}{https://huggingface.co/google/electra-base-discriminator}} are used to extract $768$ features.

\begin{table*}[!th]
 	\caption{\ac{SER} comparison of linguistic features on the \ac{IEMOCAP} corpus (Chance level: $25.0\,\%$ UAR). Every text feature extractor is tested with different \ac{ASR} systems as input as well as with the gold standard (\textbf{GS}). \textbf{UAR}: Unweighted Average Recall. \textbf{QN}: QuartzNet. \textbf{TT}: Transformer Transducer. \textbf{W2V}: wav2vec. \textbf{LM}: Language model.
 	}
	\label{tab:Baselines}
\centering
  \resizebox{1.0\textwidth}{!}{
\begin{tabular}{lrrrrrrrrrrrrrr} 
    \toprule \relax
    [UAR \%] & \multicolumn{2}{c}{\bf GS Text }  & \multicolumn{2}{c}{\bf ASR QN }  & \multicolumn{2}{c}{\bf ASR QN-LM } & \multicolumn{2}{c}{\bf ASR TT } & \multicolumn{2}{c}{\bf ASR TT-LM } & \multicolumn{2}{c}{\bf ASR W2V } & \multicolumn{2}{c}{\bf ASR W2V-LM }\\ \midrule 
    Network & Dev & Test & Dev & Test & Dev & Test & Dev & Test & Dev & Test & Dev & Test & Dev & Test \\ \midrule 
    
    \textsc{DeepMoji}       & \textit{61.7} & \textit{63.0} & 52.2 & 49.6 & 52.3 & 48.6 & 53.8 & 54.0 & 53.8 & 45.2 & \textbf{57.4} & \textbf{58.0} & 56.3 & 55.2 \\
    \textsc{BERT Base}      & \textit{57.6} & \textit{58.0} & 47.1 & 45.9 & 47.5 & 45.0 & 50.1 & 50.9 & 49.1 & 50.7 & 51.6 & \textbf{55.2} & \textbf{53.9} & 55.1 \\
    \textsc{ALBERT Base}    & \textit{47.9} & \textit{52.7} & 38.1 & 40.3 & 40.8 & 42.4 & 42.2 & 43.1 & 41.4 & 44.3 & 42.7 & 46.3 & \textbf{44.3} & \textbf{49.2} \\
    \textsc{ELECTRA Base}   & \textit{56.9} & \textit{56.2} & 44.2 & 43.1 & 46.7 & 43.2 & 48.2 & 45.5 & 47.1 & 45.6 & 52.6 & 49.7 & \textbf{53.3} & \textbf{49.9} \\
\bottomrule
\end{tabular}
}
\end{table*}

\subsection{Automatic Speech Recognition}
\label{ssec:asr}
To obtain text encodings from audio waveforms, we employ several pre-trained \ac{ASR} systems: a system based on \ac{QN}, streaming \ac{TT} and \ac{W2V}. 
\acf{TT}~\cite{transformerTransducer2020} is an end-to-end model trained with RNN-Transducer (RNN-T) loss \cite{rnnt2012}.
Its encoder implementation entails Transformer blocks with multi-headed self-attention masking future context making the network suitable for stream audio processing. The label encoder of this architecture can be interpreted as a small built-in internal \ac{LM} as it takes the previous predicted output label as input. The joint network combines audio and label encoder outputs and passes them to the the final softmax. Our solution is trained on LibriSpeech~\cite{librispeech2015}, CommonVoice~\cite{commonvoice2019}, and Tedlium~\cite{tedlium2012}. Additionally, we utilise our internal audio sets with various eastern accents for model fine-tuning. In total, about $4\,000$ hours of speech are used for \ac{TT} training. For tokenisation, we use the \ac{BPE}~\cite{sennrich-etal-2016-neural} sentence-piece model with a vocabulary size of $4\,096$ items. Despite the fact that the system has an internal \ac{LM}, we additionally evaluate our model in combination with an external \ac{LM} based on the transformer architecture. The \ac{LM} is trained on $30$ gigabytes of corpora that include wiki texts and books. Furthermore, cold fusion is used to connect the outputs of the \ac{LM} to the external \ac{LM} with the lambda parameter set to $0.2$.

\acf{QN}~\cite{quartznet2020} is a \ac{CTC}~\cite{ctc2006} loss based model composed of blocks with separable convolutions and residual connections between them, with a fully connected decoder at the end. The model has fewer parameters than \ac{TT} while still showing near state-of-the-art accuracy. In our experiments, first, a pre-trained configuration with $15$ blocks and $5$ sub-blocks in each block is used. Subsequently, we fine-tune it on the dataset with British accents and recordings generated by a \ac{TTS} system. For training the \ac{QN} model, we employ around $2\,000$ hours of internal and public datasets. Unlike \ac{TT}, we set up the configuration to predict graphemes. By default, the \ac{CTC} loss does not consider the use of a built-in \ac{LM}. We use a 4-gram statistical language model learnt on Gigaword~\cite{gigaword2012} and fuse it with \ac{QN} during the inference decoding.

The \acf{W2V}~\cite{wav2vec2020} system is a new framework for self-supervised training. The model can be broken down into three parts: 
i) a feature encoder, which represents speech waveforms in latent states that concurrently goes further to a contextualised representation part, ii) a quantisation module, iii) a contextualised representation joined with Transformers learns the relative positional information in the latent states. A quantization module represents the infinite output of a feature encoder to a discrete set via product quantization. The framework exploits the \ac{CTC} loss for training. We use a pre-trained large model from the official repository\footnote{\href{https://github.com/pytorch/fairseq/tree/master/examples/wav2vec}{https://github.com/pytorch/fairseq/tree/master/examples/wav2vec}}. First, we choose the checkpoint obtained by the training on $60\,k$ hours of LibriVox, subsequently, we finetune it with additional $960$ hours of LibriSpeech. For the external language model, we fuse \ac{W2V}, which is the same as for \ac{QN}.

\section{Experiments}
\label{sec:experiments}

\subsection{Dataset}
\label{ssec:dataset}
The \acf{IEMOCAP} dataset~\cite{busso2008iemocap} is an English emotion dataset that comes with both textual transcriptions and raw audio information. The dataset contains scripted and improvised dialogues between $5$ female and $5$ male speakers.
In order to be consistent with previous research with \ac{IEMOCAP}, we choose the main emotions happiness (fused with excitement), sadness, anger, and neutral.
The choice of emotions results in $5\,531$ utterances totaling $7.0$ hours of audio data. In literature, there is no agreement on the partitioning of the dataset. In our experiments,
we split the \ac{IEMOCAP} dataset speaker-independently into session $1-3$ for training, session $4$ for development, and session $5$ for testing.

LibriSpeech~\cite{librispeech2015} is a publicly available and popular dataset for speech recognition system experiments and evaluations. It includes transcriptions for around $960$ hours of public domain audio books dictated by many speakers. For our experiments, we use test-clean part which is about $4.5$ hours of audio with $20$ male and $20$ female speakers.

\subsection{Automatic Speech Recognition}
\label{ssec:asr_experiments}
We evaluate each of our models on LibriSpeech test-clean with and without the external \ac{LM} to compare the accuracy. We measure \acf{WER} and \acf{CER}. As two models are adopted for different accents, state-of-the-art results for the chosen data are not expected. On the contrary, the \ac{W2V} trained on data with the same distribution as LibriSpeech and setups with this model show more precise predictions for both datasets. The evaluation on test-clean compares the accuracy of models on public and well-known data. Our measurements for \ac{IEMOCAP} and LibriSpeech test-clean are demonstrated in~\Cref{tab:ASR}.

\begin{table}[bh!]
 	\caption{Evaluation results of \acp{ASR} on LibriSpeech test-clean and \ac{IEMOCAP} waveforms. \textbf{WER}: Word error rate. \textbf{CER}: Character error rate.}
	\label{tab:ASR}
\centering
  \resizebox{1.0\columnwidth}{!}{
\begin{tabular}{lrrrr}
    \toprule
    \multicolumn{1}{l}{\bf SetUp} &\multicolumn{2}{c}{\bf LibriSpeech} & \multicolumn{2}{c}{\bf IEMOCAP } \\ \midrule 
    &   WER [\%]  &   CER [\%] & WER [\%] & CER [\%] \\ 
    \midrule
    \textsc{QN}             & 15.69     &  7.26     & 41.21     & 25.43 \\
    \textsc{QN + LM}        & 12.98     &  7.35     & 42.14     & 31.21 \\
    \textsc{TT}             & 4.98      &  1.89     & 31.81     & 22.30 \\
    \textsc{TT + LM}        & 4.93      &  1.86     & 31.47     & 22.08 \\
    \textsc{W2V}            & \bf2.16   &  \bf0.57  & 25.71     & 13.56 \\
    \textsc{W2V + LM}       & 2.24      &  0.63     & \bf22.23  & \bf13.37 \\
\bottomrule
\end{tabular}
}  
\end{table}

\begin{table}[!th]
 	\caption{SER Results of audio features on \ac{IEMOCAP}. The best resulting model of \textsc{DeepSpectrum} is a \textsc{DenseNet201} with $128$ Mel bins and the \textit{viridis} colour map.}
	\label{tab:audio_only_results}
\centering 
\begin{tabular}{lrr}
    \toprule
    {[UAR \%]} &   \textbf{Dev}  &   \textbf{Test }   \\
    \midrule                                             
    \textsc{openSMILE} (ComParE\_2016) & \bf57.8  &  58.5    \\
    \midrule
    \textsc{openXBOW} (\(N=2\,000\))            & 55.7  &  59.1    \\ 
    \midrule
    \textsc{DeepSpectrum}    & 53.2  &  \bf59.8    \\

    \midrule
    \textsc{auDeep} (\(X=-60\,dB\))             & 55.0  &  53.3    \\
\bottomrule
\end{tabular}
\end{table}

\begin{figure*}[th!]
\begin{center}
\includegraphics[width=1\textwidth]{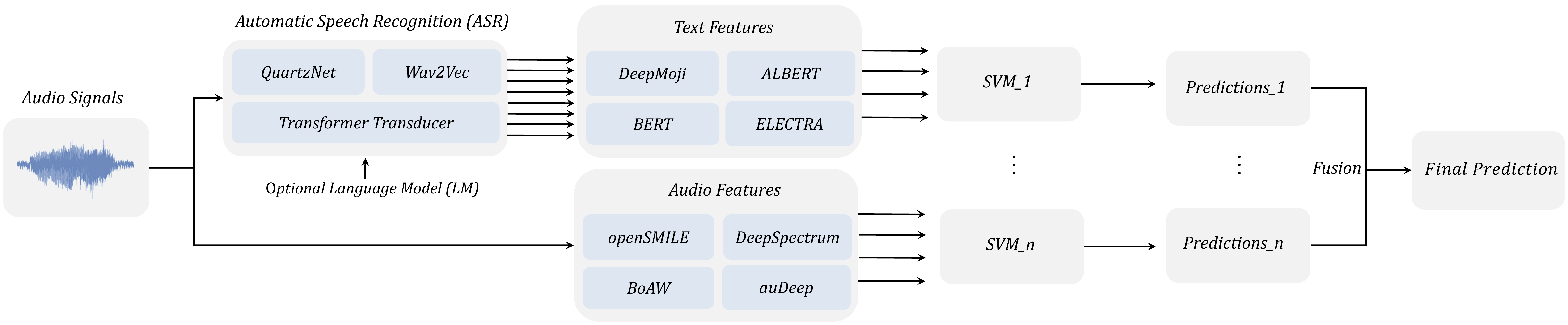}
\end{center}
\caption{A general overview of our emotion recognition by multi-modal fusion of different trained \acfp{SVM}.}
\label{fig:system_overview}
\end{figure*}

\subsection{Late Fusion}
\label{ssec:latefusion}
We apply late fusion in form of majority voting on the basis of \ac{SVM} predictions on the individual feature sets (cf.~\Cref{fig:system_overview}).
We evaluate all combinations of at least three feature sets from the audio, \ac{GS} human transcribed spoken content, and \ac{ASR}-based spoken content transcription. \Cref{tab:Fusion} summarises our results.
Taking into account a high number of feature set combinations, we restrict our evaluation of the \ac{ASR}-based text features to the \ac{ASR} system with the best overall baseline performance which is \ac{W2V} with \ac{LM}.
Feature set combinations, based on the \ac{GS} text only, clearly outperform those, which are based on \ac{ASR}-generated text. Considering the fact that this result is in line with performance of the individual feature sets (showed in~\Cref{tab:Baselines}), most likely, this behaviour is caused by the \ac{WER} of the \ac{ASR} system. 
However, when combining audio and text features, both \ac{ASR}-based and \ac{GS}-based feature sets lead to a similar performance, being higher than any individual information stream, both on average and for the best performance.
Combinations of \ac{ASR}-based and \ac{GS}-based features show no improvement in the average \ac{UAR} compared to the \ac{GS} feature sets, most likely due to the similarity of the features. 
When considering all combinations of available feature sets the average and best performance can be further increased, which can most reasonably be explained by the vast number of combinations. 
Several combinations consisting of audio, \textsc{\ac{GS}-Text} and \textsc{\ac{ASR}-Text} features -- including the combination \textsc{GS-BERT}, \textsc{GS-DeepMoji}, \textsc{ASR-BERT}, \textsc{ASR-DeepMoji, ASR-ELECTRA, auDeep, DeepSpectrum, openSMILE, BoAW} -- achieve the best observed performance of $73.6\,\%$ \ac{UAR} on the development set and a corresponding \ac{UAR} of $73.8\,\%$ on the test set.

\begin{table}[th!]
 	\caption{Results of the majority voting late fusion.
 The possible number of feature set combinations (\#), as well as the mean UAR and standard deviation are reported. We further provide the UAR of the best performing feature set combination on the development set, as well as the corresponding performance of said combination on the test set.
 }
	\label{tab:Fusion}
\centering
  \resizebox{1.0\columnwidth}{!}{
\begin{tabular}{lrrrr}
    \toprule
    {[UAR \%]} &   \textbf{\#}  &   \textbf{Mean Dev} & \textbf{Max Dev}  & \textbf{Test} \\ 
    \midrule
    \textsc{Audio}          & 5     &  $59.9 \pm 0.1$ & 60.8 & 63.3 \\
    \textsc{GS-Text}           & 5     &  $60.3 \pm 1.5$ & 61.7 & 63.0 \\
    \textsc{ASR-Text}            & 5     &  $54.6 \pm 1.0$ & 55.8 & 56.2 \\
    \textsc{Audio + GS-Text}   & 219   &  $65.0 \pm 3.5$ & 71.0 & 69.9 \\
    \textsc{Audio + ASR-Text}    & 219   &  $63.4 \pm 3.2$ & 69.5 & 70.5 \\
    \textsc{ASR-Text + GS-Text}     & 219   &  $58.8 \pm 3.6$ & 63.3 & 64.8 \\
    \textsc{All Systems}    & 4017  &  \textbf{66.2} $\pm\ 3.8$ & \textbf{73.6} & \textbf{73.8} \\
\bottomrule
\end{tabular}
}
\end{table}

\section{Discussion}
\label{sec:discussion}
When comparing the performance of different ASR systems in Table~\ref{tab:ASR} and Table~\ref{tab:Baselines}, a correlation between low WER values and high UAR values becomes obvious. Accordingly the Gold Standard System using human annotations, which is considered to have a much lower WER than any of the ASR systems, clearly achieves the highest UAR. A similar effect has previously been observed in~\cite{metze2010}, however, utilising a --~from today's point of view~-- outdated ASR systems with a much more limited vocabulary size.
Table~\ref{tab:Fusion} suggests that a higher number of considered feature sets leads to a higher \ac{UAR} on average. 
This effect is known in general, however, it should be noted that the pairwise dependence of classifiers plays a considerable role in such a late fusion system~\cite{Kuncheva03-LMV}. A pairwise dependence of ASR-based and GS-based feature sets could therefore explain why combinations of both sets perform worse than combinations which combine either ASR-based or GS-based feature sets with audio-based feature sets. 
Assuming a high dependence between ASR-based and GS-based feature sets would further suggest that a well-suited weighted fusion method combining only audio-based and ASR-based features might further increase results towards the best-performing configuration introduced in~\ref{ssec:latefusion}, which combines two instances of BERT and DeepMoji features.

\section{Conclusions}
\label{sec:conclusions}
In this paper, we presented current ASR systems to create transcriptions for the linguistic SER. Without adapting the ASR systems to the target database IEMOCAP, we were able to achieve state-of-the-art results by fusing acoustic and linguistic information. We further observed that higher WERs on the ASR systems lead to higher UAR values for emotion recognition.

For future work, the number of feature sets and the respective feature set sizes can be reduced in order to increase computational efficiency. Furthermore, evaluation could be performed on more natural or in-the-wild databases. We mainly chose IEMOCAP as it is widely established and thereby suitable for comparison with state-of-the-art approaches. Moreover, IEMOCAP contains transcriptions making it easier to evaluate the impact of WER achieved by the ASR on the final emotion classification. Finally, the fusion with ASR could be implemented on the levels of the embeddings instead of the text.

\section{Acknowledgements}
This research was partly supported by the Affective Computing \& HCI Innovation Research Lab between Huawei Technologies and University of Augsburg. We acknowledge funding from Deutsche Forschungsgemeinschaft (DFG) under grant agreement No.\ 421613952 (ParaStiChaD), and Zentrales Innovationsprogramm Mittelstand (ZIM) under grant agreement No.\ 16KN069455 (KIRun). The work of Artem Sokolov is partially supported by RSF (Russian Science Foundation) grant 20-71-10010.

\clearpage

\bibliographystyle{IEEEtran}

\bibliography{mybib}

\begin{acronym}
\acro{CNN}[CNN]{Convolutional Neural Network}
\acrodefplural{CNN}[CNNs]{Convolutional Neural Networks}
\acro{RNN}[RNN]{Recurrent Neural Network}
\acrodefplural{RNN}[RNNs]{Recurrent Neural Networks}
\acro{LSTM}[LSTM]{long short-term memory}
\acro{MLP}[MLP]{multi-layer perceptron}
\acrodefplural{MLP}[MLPs]{multi-layer perceptrons}
\acro{HDF}[HDF]{Hierarchical Data Format}
\acro{STFT}[STFT]{Short-Time Fourier Transform}
\acrodefplural{STFT}[STFTs]{Short-Time Fourier Transforms}
\acro{SVM}[SVM]{Support Vector Machine}
\acrodefplural{SVM}[SVMs]{Support Vector Machines}
\acro{IEMOCAP}[IEMOCAP]{Interactive Emotional Dyadic Motion Capture}
\acro{UAR}[UAR]{Unweighted Average Recall}
\acrodefplural{UAR}[UARs]{Unweighted Average Recall}
\acro{TF}[TF]{TensorFlow}
\acro{ComParE}[ComParE]{Computational Paralinguistics Challenge}
\acrodefplural{ComParE}[ComParE]{Computational Paralinguistics Challenges}
\acro{WER}[WER]{Word error rate}
\acro{CER}[CER]{character error rate}
\acro{SER}[SER]{speech emotion recognition}
\acro{ASR}[ASR]{automatic speech recognition}
\acro{LLD}[LLD]{low-level descriptor}
\acrodefplural{LLD}[LLDs]{low-level descriptors}
\acro{BoAW}[BoAW]{Bag-of-Audio-Words}
\acrodefplural{BoAW}[BoAWs]{Bag-of-Audio-Words}
\acro{TT}[TT]{Transformer Transducer}
\acro{QN}[QN]{QuartzNet}
\acro{W2V}[W2V]{wav2vec}
\acro{LM}[LM]{language model}
\acro{BPE}[BPE]{Byte-Pair Encoding}
\acro{CTC}[CTC]{Connectionist Temporal Classification}
\acro{TTS}[TTS]{text-to-speech}
\acro{GS}[GS]{Gold Standard}
\acro{BERT}[BERT]{Bidirectional Encoder Representations from Transformers}
\end{acronym}

\end{document}